\documentclass[fleqn,12pt,twoside]{article}
\usepackage{espcrc1}

\usepackage{graphicx}
\usepackage{color}
\usepackage{psfrag}
\usepackage{epsfig}

\usepackage{mymargins}

\def\gsi{\raise0.3ex\hbox{$>$\kern-0.75em\raise-1.1ex\hbox{$\sim$}}}
\newcommand{\gsim} {\mathop{\gsi}}
\newcommand{\be} {\begin{equation}}
\newcommand{\ee} {\end{equation}}

\title{The baryon static potential from lattice QCD}  

\author{Ph. de Forcrand\address[MCSD]{Institute for Theoretical Physics,
        ETH, CH-8093 Z\"urich, Switzerland, and \\
        CERN, Department of Physics, TH Division, CH-1211 Geneva 23, Switzerland}
        and
        O. Jahn\address{Center for Theoretical Physics, MIT, 
        Cambridge, MA 02139, USA} }
       
\begin{document}

\maketitle

\begin{abstract}
Lattice QCD simulations offer the possibility of determining the
potential between three static quarks from first principles.
We review the status of such simulations, and the relative standing
of the two theoretical proposals for the baryonic potential:
the Delta law (sum of two-body terms) and the Y law (length of
three flux strings joined together at a junction).
We also present new results on the leading L\"uscher-like corrections
to the asymptotic linear potential.
\end{abstract}

\section{INTRODUCTION}

The potential between 3 static color charges can help predict the masses of
baryons made of 3 heavy, non-relativistic quarks. It also reveals by which
mechanism the color interaction confines the 3 quarks into a color singlet.
In the $q\bar{q}$, mesonic case, the 2 charges are confined by the formation
of a string of flux joining them. The string energy grows in proportion to
its length, which produces linear confinement. Moreover, the worldsheet spanned
by this string fluctuates, and the Casimir energy of such Gaussian 
fluctuations is expressed by the Dedekind $\eta$-function, which 
generates a $\frac{\pi (d-2)}{24} \frac{1}{r}$ correction to
the linear potential in $d$ dimensions, the so-called L\"uscher correction \cite{luschterm}.
This correction is universal, because it does not depend on the underlying 
linearly confining theory.

The numerical study of linear confinement in the baryonic case has been the
object of old \cite{old} and new \cite{us1,us2,us3,them1,them2} lattice simulations. 
The phenomenological
question is: is confinement accompanied by the formation of strings of flux
as in the $q\bar{q}$ case? and if so, is the string tension the same?
The answer is yes to both questions, and we review the evidence below in
Sec.~II. When the separation between any 2 quarks is large, 3 strings of color 
flux form, which meet at a junction. Their worldsheet forms a 3-bladed surface,
whose Casimir energy produces a L\"uscher-like $\frac{1}{r}$ correction,
which depends on the geometry of the $qqq$ triangle but not on the confining
theory. This correction can be computed analytically \cite{us3}, and we compare it with
numerical simulations in Sec.~III.

\section{CHANGING FROM $\Delta$-LAW TO $Y$-LAW WITH INCREASING DISTANCE}

At short distance, perturbation theory applies. One-gluon exchange between 
2 quarks $i$ and $j$ of a baryon at distance $r_{ij}$ produces an attractive
force, which is 1/2 that between $q$ and $\bar{q}$ at the same distance,
because of the different color index contractions. Therefore, the baryonic
potential at short distances obeys the so-called Delta law:
\be
\label{Delta}
V_{qqq}(\vec{r}_1,\vec{r}_2,\vec{r}_3) \approx \frac{1}{2} \sum_{i<j} V_{q\bar{q}}(r_{ij})
\ee
In this case, only 2-body $qq$ forces are at work. If this ansatz persists
at large distances, then the potential will grow linearly as $\frac{1}{2}
\sigma_{q\bar{q}} (r_{12} + r_{23} + r_{31})$, hence the name $\Delta$-law.

At large distances, if 3 flux strings form, with energy proportional to 
their length, they will meet at a junction $x_J$ which in the groundstate will localize
at the Steiner point which minimizes the total
string length $L_Y = \min_{x_J} \sum_{i=1}^3 |\vec{r}_i - \vec{x}_J|$. The potential
will grow as the $Y$-law:
\be
\label{Y}
V_{qqq}(\vec{r}_1,\vec{r}_2,\vec{r}_3) \approx \sigma_{q\bar{q}} L_Y
\ee

\begin{figure}[t]
\begin{minipage}[t]{76mm}
\hbox{\includegraphics[width=1.0\textwidth,clip,trim=0 0 0 278]{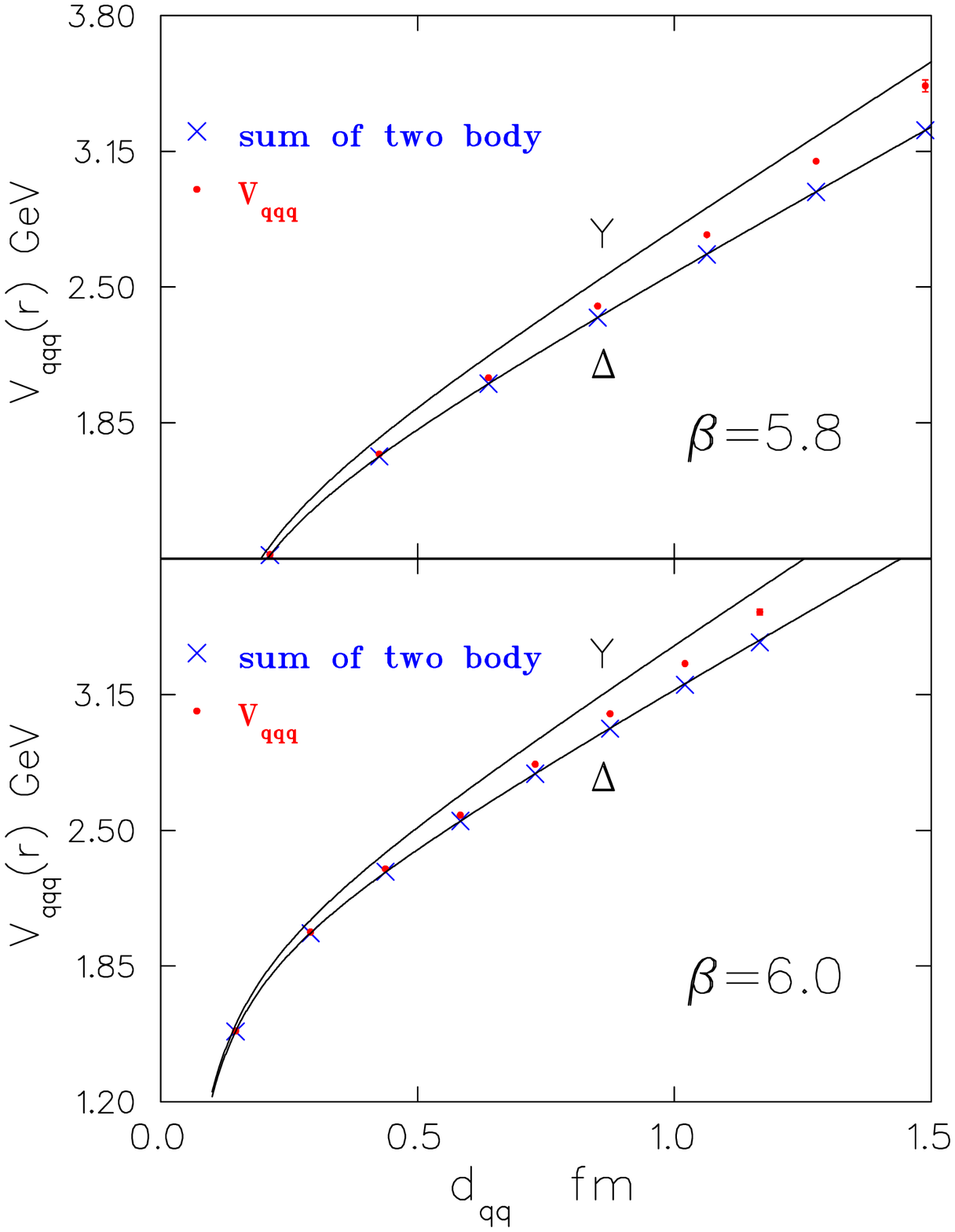}}
\vspace*{-10mm}
\caption{Static baryon potential for equilateral geometry, showing agreement
with $\Delta$-law at short distances and departure for $d_{qq} \gsim 0.8$ fm.
From \cite{us2}.}
\label{fig:largenenough}
\end{minipage}
\hspace*{4mm}
\makeatletter
\def\fnum@figure{Table I}
\makeatother
\begin{minipage}[t]{70mm}
\label{table:1}
\vspace*{-53mm}
\hspace*{-2mm}
\begin{tabular}{ccccc}
\hline
$\beta$      &  $a$(fm)  & $\sigma_{q \bar{q}}$  & $\sigma_{qqq}$ & $1 - \frac{\sigma_{qqq}}{\sigma_{q \bar{q}}}$ \\
\hline
5.7          &  0.19     & 0.1603            & 0.1556     & 3\%     \\
             &           & ~~~~~48            & ~~~~~24     &         \\
\hline
5.8          &  0.14     & 0.1080            & 0.1031      & 4.5\%   \\
             &           & ~~~~~28            & ~~~~~~~6     &         \\
\hline
6.0          &  0.10     & 0.500              & 0.467       & 6.6\%   \\
             &           & ~~~~~~7            & ~~~~~4     &         \\
\hline
\end{tabular}
\vspace*{-5mm}
\caption{Results from \cite{them2}, demonstrating departure from $Y$-law
         at short distances (high $\beta$). At short distances, the effective
$\sigma_{qqq}$ is reduced, indicating that the potential grows more slowly,
as expected from the $\Delta$-law.}
\end{minipage}
\end{figure}
\addtocounter{figure}{-1}

Because the difference between $L_Y$ and $\frac{L_\Delta}{2}$ is at most
$\sim 15\%$, it takes good numerical accuracy to distinguish between the
two ans\"atze. Moreover, the $\Delta$-law holds at short distances, so 
the question really is: at which distance does the $Y$-law take over,
if it does?
Two groups have addressed this question with the required accuracy. \\
$\bullet$ Ref.\cite{us1} points out that no significant deviation from the 
$\Delta$-law is observable up to $qq$ separations of about 0.8 fm. In the
follow-up Ref.\cite{us2}, several technical refinements, among them a 
variational basis of junction locations, allow to uncover clear deviations
from the $\Delta$-law, consistent with the $Y$-law. Fig.~1, taken from 
\enlargethispage*{\baselineskip}
Ref.\cite{us2}, shows the difference between the measured $V_{qqq}$ and the
measured $\Delta$-law eq.(\ref{Delta}). Note that no fitting is 
involved.
At distances larger than $\sim 0.8$ fm, the slope of $V_{qqq}$ (i.e. the
baryonic force) becomes consistent with that of the $Y$-law eq.(\ref{Y}). \\
$\bullet$ The other group \cite{them1,them2} obtains similar numerical results, but analyzes
them differently. The data are fitted with a $\Delta$- or $Y$-ansatz, and
the latter works best. In this fit, the baryonic string tension $\sigma_{qqq}$
is a fitting parameter, allowed to differ from the measured $\sigma_{q\bar{q}}$.
A selection of the results of Ref.\cite{them2} is shown in Table I, illustrating
that at higher $\beta$, where the lattice is fine and small distances 
dominate the fit, $\sigma_{qqq}$ comes out systematically smaller than 
$\sigma_{q\bar{q}}$. 
Equivalently, if one would enforce instead $\sigma_{qqq} = \sigma_{q\bar{q}}$, 
the effective string length is smaller
than $L_Y$, increasingly so at shorter distances (higher $\beta$), as
expected from the transition to a $\Delta$-law at short distance.
So the results of Ref.\cite{them2} are simply inconsistent with their simplifying
statement that ``the Coulomb plus Y-type linear potential is accurate at the 1\% level''.
The $\Delta$- to $Y$-law transition is gradual, and occurs around 
$qq$ separations ${\cal O}$(0.8 fm). The string tension 
$\sigma_{qqq} = \sigma_{q\bar{q}}$ defines the unique correlation length
emerging from the gluon dynamics. The dominance of the $Y$-law occurs
for strings of length $\sim 0.5$ fm or more. At shorter distances,
the ``string'' is as fat as it is long, and the flux picture underlying
the $Y$-law breaks down. This is similar to the ${q\bar{q}}$ case.
However, it indicates that a flux tube description  of a baryon, as e.g.
in \cite{Page}, may be a mediocre approximation.

Baryonic flux strings have been exhibited, in quenched and full QCD after
Abelian projection \cite{Ichie}, and in a gauge-invariant way after 
smearing \cite{Leinweber}.

\section{BARYONIC L\"USCHER-LIKE CORRECTION}

At large distances, where the string description applies, the worldsheet of the
three strings describes a 3-bladed surface. Assigning to such a surface an
action proportional to its area, one can analytically integrate over small
Gaussian fluctuations, much like in the $q \bar{q}$ case. The essential
difference is in the boundary conditions: each of the 3 sheets has one fixed
($q$) and one fluctuating (junction) boundary. At the junction, continuity
and balance of forces cause a mixed Dirichlet-Neumann condition \cite{tH}. Summation
over all eigenenergies results in a L\"uscher-like correction to the 
$V_{qqq}$ potential \cite{us3}:
\be
C \frac{1}{L_Y}
\ee
where $C$ depends on the geometry of the $qqq$ triangle and on the space-time
dimension. \\
$\bullet$ In $d=3$, $C$ is non-negative. Therefore, the linear asymptotic
behaviour $V_{qqq} \sim \sigma L_Y$ is approached from {\em above}, and the 
potential has an inflexion point as a function of the triangle size (for
fixed geometry). For the equilateral case, $C=0$, due to a cancellation between
the 2 types of eigenmodes: all 3 blades vibrating in phase (1 mode), and
2 blades in phase opposition with the third at rest (3 modes, of which 2 are
independent), carrying a factor -1/2 due to the different boundary conditions.\\
$\bullet$ In $d=4$, fluctuations of the junction outside the $qqq$ plane 
generate a large negative $\frac{1}{L_Y}$ correction, making $C$ negative
always, with less sensitivity to the $qqq$ geometry. \\
Fig.~2 shows $C \times \frac{24}{\pi}$ in $d=3$ and 4, as a function of the $qqq$ triangle geometry,
represented by coordinates $(L_1,L_2)$ which are the lengths of 2 blades
relative to $L_3=1$. The equilateral case is $(1,1)$, the right-angled isosceles
case is $(\sqrt{3}+1,\sqrt{3}+1)$.

\begin{figure}[t]
\begin{minipage}[t]{65mm}
\hbox{\includegraphics[height=65mm]{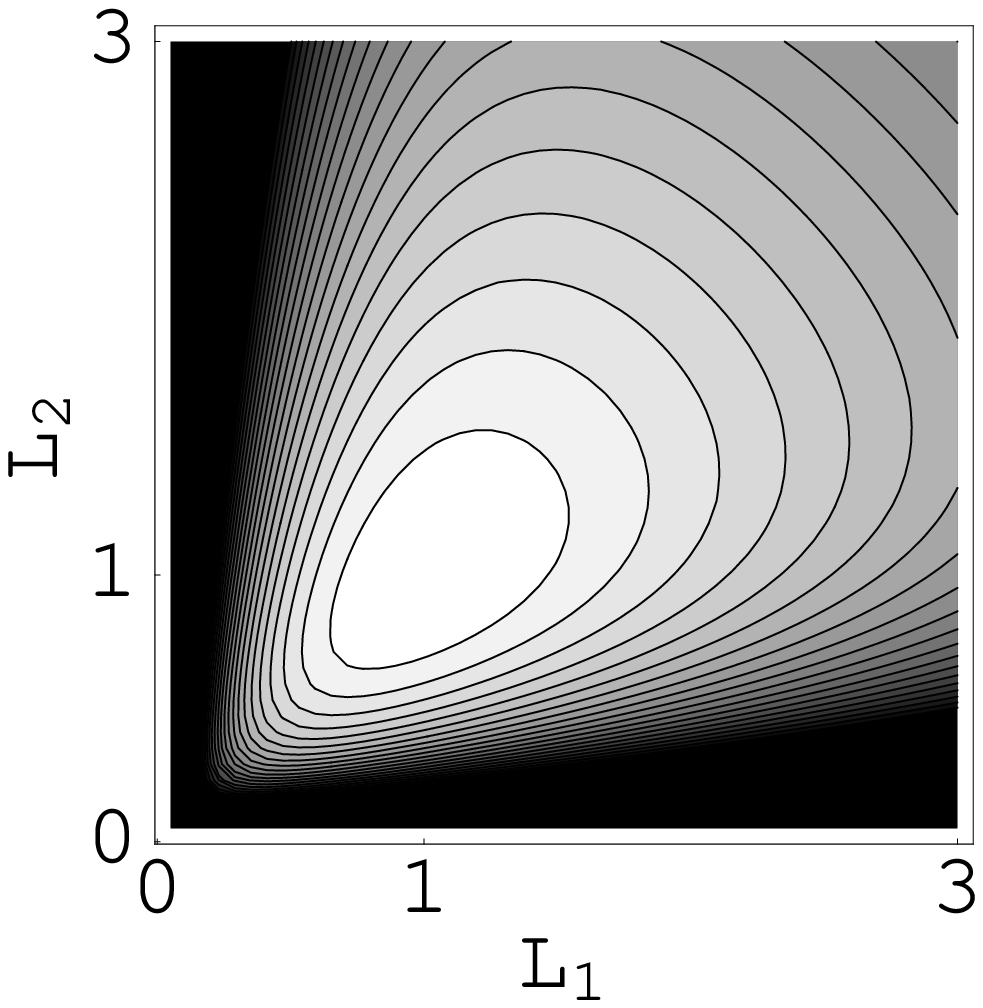}}%
\end{minipage}
\hspace{-3mm}
\begin{minipage}[t]{0mm}
\hbox{\includegraphics*[height=65mm]{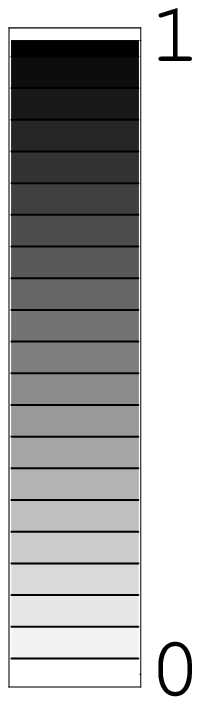}}
\end{minipage}
\begin{minipage}[t]{70mm}
\hspace{-1mm}
\hbox{\includegraphics[height=65mm]{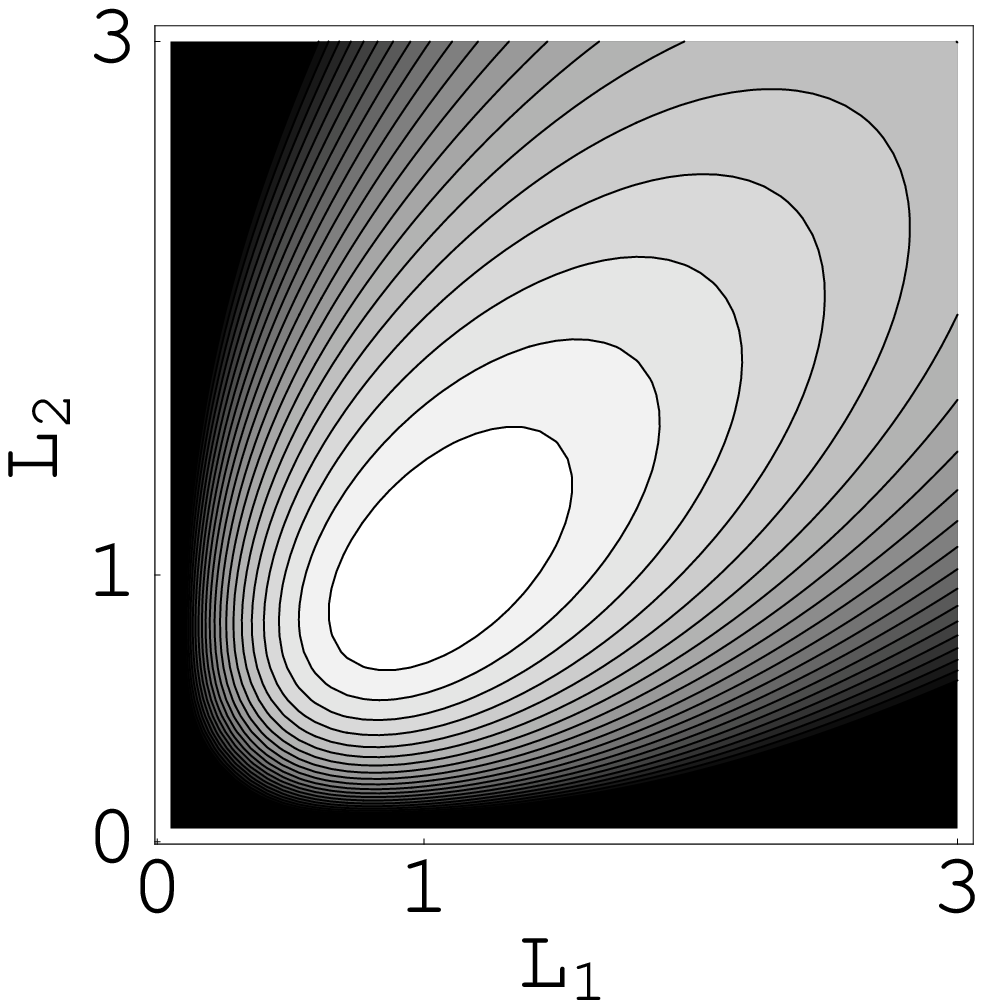}}%
\hbox{\includegraphics*[height=75mm]{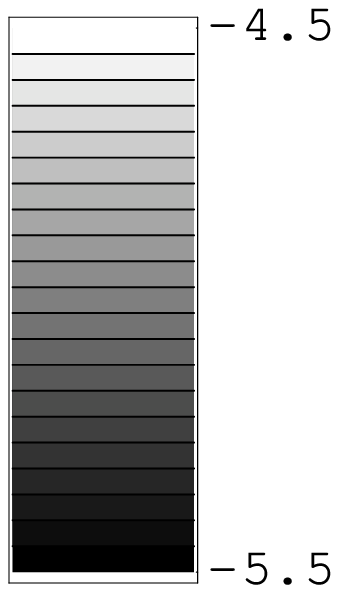}}
\end{minipage} \\
\vspace*{-10mm}
\caption{Coefficient of the L\"uscher-like term $\frac{\pi}{24} \frac{1}{L_Y}$
as a function of the 3 relative string lengths $(L_1,L_2,L_3=1)$,
in $d=3$ (left) and 4 (right). 
In $d=3$, $C > 0$, so that $V_{qqq}$ has an inflexion point.}
\end{figure}

These analytic predictions can be confronted to numerical simulations.
Since they are universal, we chose to simulate the simplest gauge theory,
$Z_3$ in $d=3$. To reach the required distances and accuracy, we first 
perform a duality transformation. A Wilson loop expectation value in the 
$Z_3$ gauge theory is equal to the free energy of an interface bounded by
this loop, in the dual 3-states Potts model. We measure the variation of
this free energy with the area of the interface by using the ``snake''
algorithm \cite{snake}, which increases the interface area one plaquette at 
a time. In fact, only one simulation is required for each elementary 
displacement of a quark as in \cite{Vettorazzo}.

\begin{figure}[h]
\vspace*{-5mm}
\begin{minipage}[h]{76mm}
\label{qqbar1}
\hbox{\includegraphics[width=1.0\textwidth]{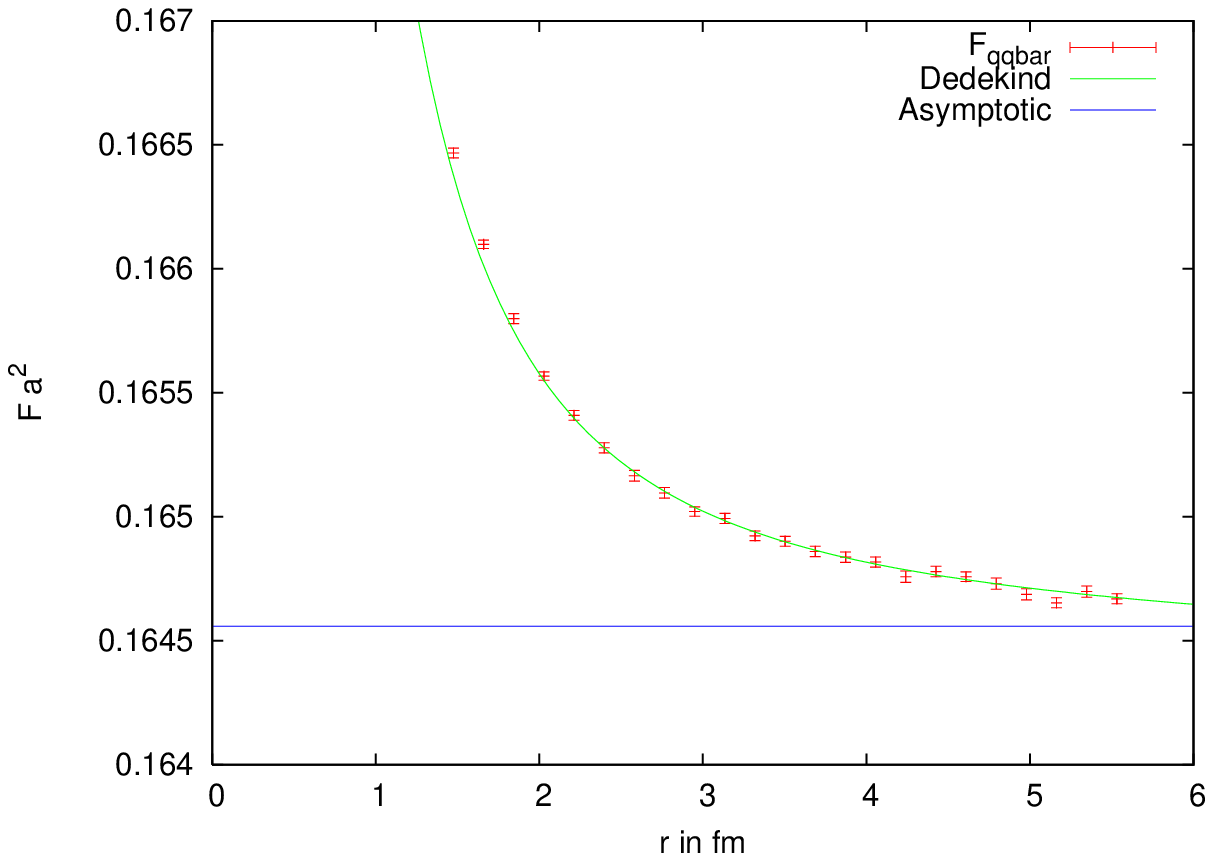}}
\vspace*{-7mm}
\end{minipage}
\hspace{\fill}
\begin{minipage}[h]{76mm}
\label{qqbar2}
\hbox{\includegraphics[width=1.0\textwidth]{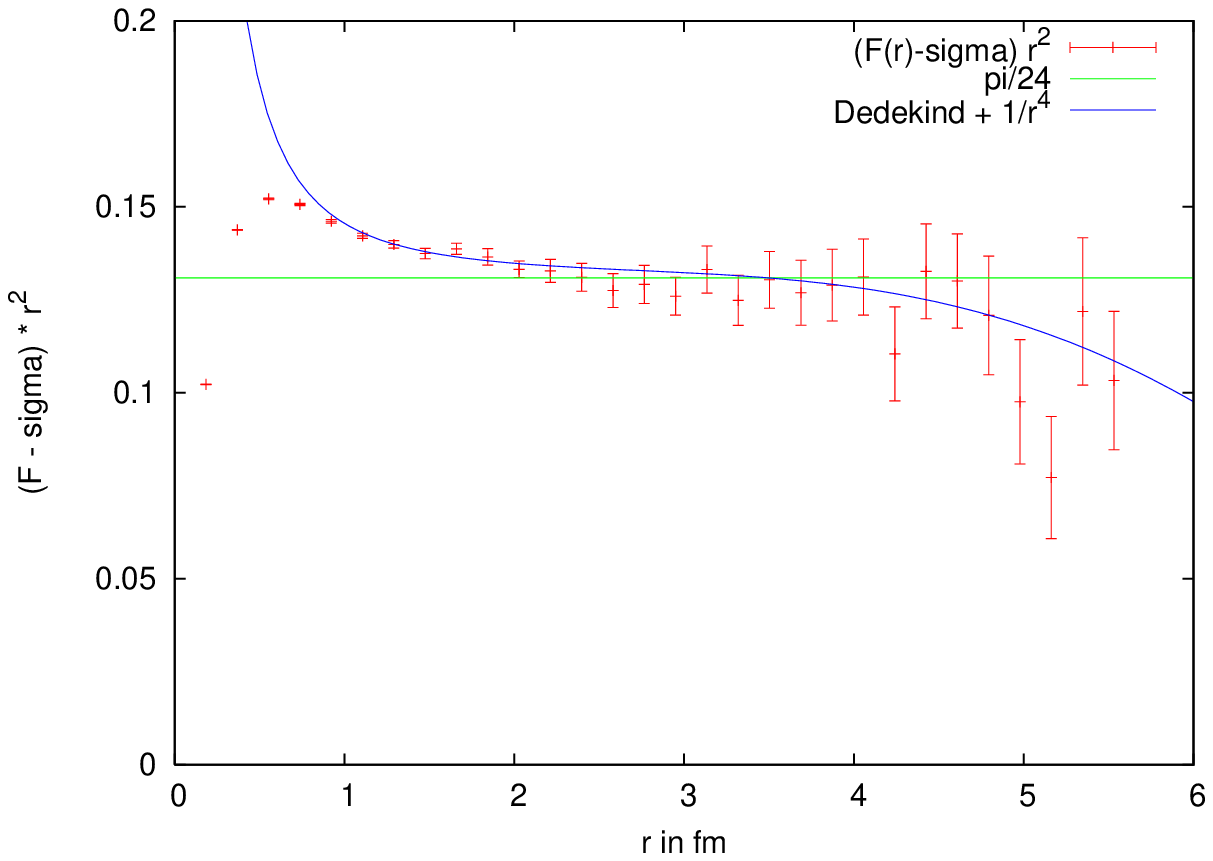}}
\vspace*{-7mm}
\end{minipage}
\caption{Mesonic ($q \bar{q}$) force in the $3d$ $Z_3$ gauge theory: force versus
distance (left) and coefficient of the L\"uscher correction (right). The solid
curve corresponds to the Dedekind function at the temperature of our study (hence the bending down),
plus a $1/r^4$ term with a fitted coefficient. The overshoot above 
$\pi/24$ has been predicted \cite{Luscher} but not yet observed in $SU(N)$ simulations.
Note the very large distances probed.}
\vspace*{-5mm}
\end{figure}

All simulations are performed in the 3-states Potts model at $\beta=0.60$
on a $48^3$ lattice, using Swendsen-Wang updates and a multishell arrangement
for variance reduction \cite{Hasenbusch}.
The crucial point of the algorithm is that a determination of the baryonic
force to a given accuracy requires computer resources {\em independent}
of the $qqq$ separation. This makes the approach extraordinarily efficient.

As an illustration, Fig.~3 shows the mesonic force $-\frac{dV_{q\bar{q}}}{dr}$
for $q\bar{q}$ distances up to $\sim 6$ fm 
(the lattice spacing $a \sim 0.18$ fm is obtained by fixing the
string tension $\sqrt{\sigma} = 440$ MeV). The corresponding L\"uscher term
indeed approaches the expected $\frac{\pi}{24} \frac{1}{r^2}$, with a
$\frac{1}{r^4}$ subleading correction of the sign predicted in \cite{Luscher}, with 
a large but $a$-dependent magnitude.

In the baryonic case, we studied 3 $qqq$ geometries: equilateral, right-angled isosceles,
and quark-diquark-like (2 quarks fixed, the third at distance $h$ along the mediatrix). In all 3 cases,
the asymptotic $Y$-law is clearly established, with $\sigma_{qqq} = \sigma_{q\bar{q}}$
(compare the force in Fig.~3 (left) and Fig.~4 (left)).
But the approach to asymptotia is very geometry-dependent, as seen in Fig.~4 (left).
It is fastest in the equilateral case, as predicted since the leading correction vanishes
for this geometry ($C=0$). Indeed, our quasi-equilateral data are well described by a force
$(\sigma - c_1/L_Y^2 + c_2/L_Y^4)$, and the fitted value of $c_1$ is consistent with the
string prediction (Fig.~4, middle). This is also true for the right-angled isosceles case,
but the subleading correction is very large. For the quark-diquark-like case, the geometry changes
as the third quark moves. The string prediction is shown by the solid curve in Fig.~4 (right),
together with the mesonic value $\pi/24$ corresponding to the quark-diquark asymptotic limit.
The data are completely consistent with the string prediction. 
Note however the large quark separations necessary to recover good agreement in all cases.

Therefore, although subleading corrections turn out to be sizable, our numerical study confirms
quantitatively the analytic calculation of the leading, universal $1/L_Y$ correction to the $Y$-law.

\begin{figure}[h!]
\vspace*{-8mm}
\hspace*{-4mm}
\begin{minipage}[h!]{74mm}
\label{qqq1}
\hbox{\includegraphics*[width=0.70\textwidth]{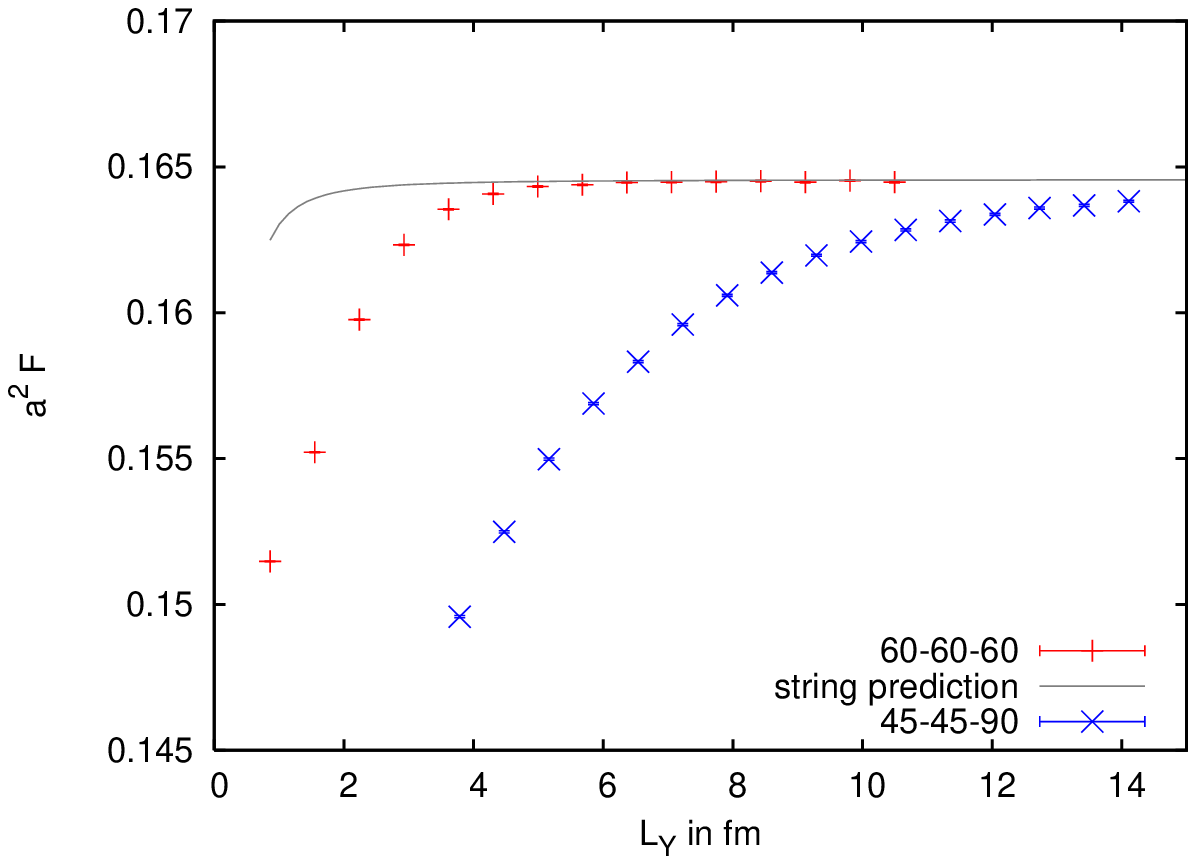}}
\vspace*{-7mm}
\end{minipage}
\begin{minipage}[h!]{74mm}
\label{qqq2}
\hbox{\includegraphics*[width=0.70\textwidth]{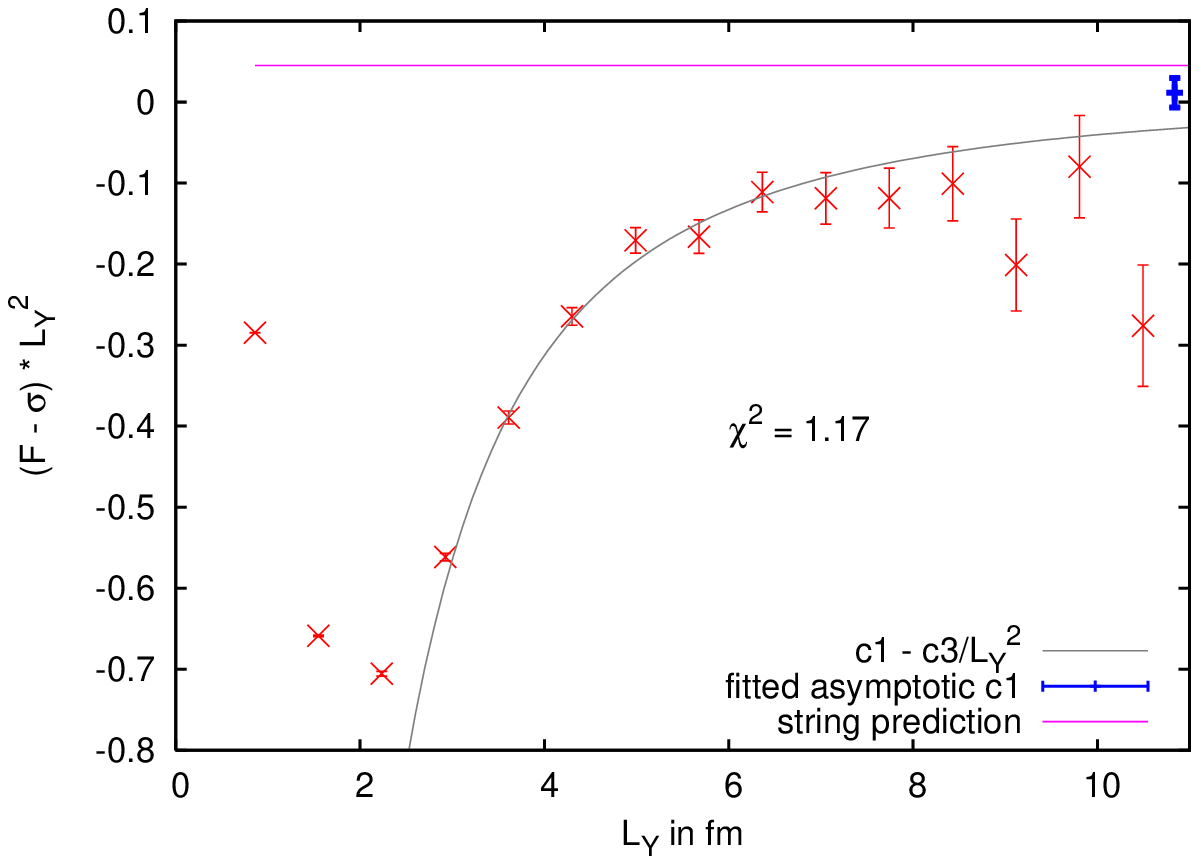}}
\vspace*{-7mm}
\end{minipage}
\begin{minipage}[h!]{74mm}
\label{qqq3}
\hbox{\includegraphics*[width=0.70\textwidth]{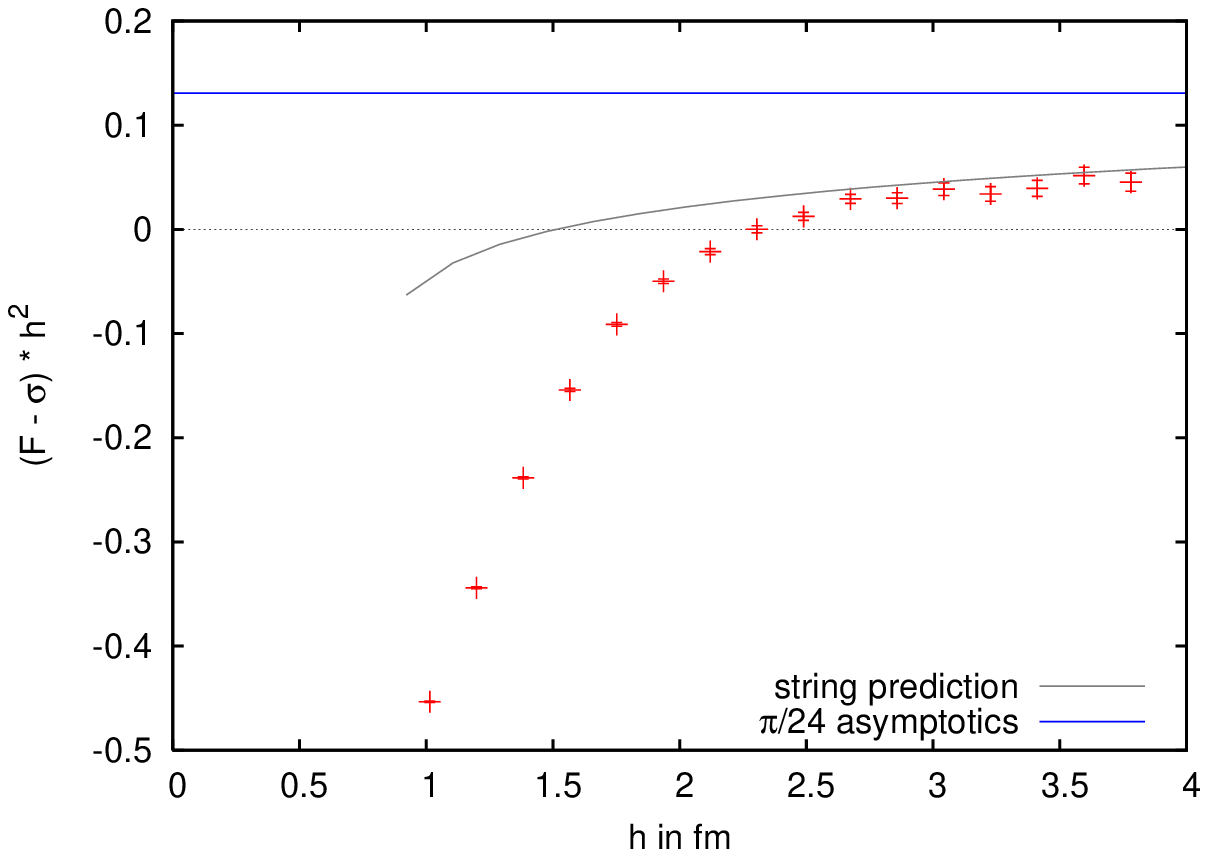}}
\vspace*{-7mm}
\end{minipage}
\caption{Baryonic ($qqq$) force and L\"uscher term in the $3d$ $Z_3$ gauge theory. 
Left: force versus distance for equilateral and 45-45-90 degree geometries.
Middle: the coefficient $C$ (equilateral case) of the L\"uscher term is consistent with
the string prediction $C \approx 0$ after including a subleading $1/r^4$ correction. 
Right: $C$ (quark-diquark case) is consistent with the string prediction and approaches $\pi/24$
in the diquark limit.
}
\vspace*{-11mm}
\end{figure}

\end{document}